# Low-Rank Structured Clutter Covariance Matrix Estimation for Airborne STAP Radar

Tao Zhang, Haifang Zheng, Qijun Luo
Tianjin Key Laboratory for Advanced Signal Processing
Civil Aviation University of China
Tianjin, China
t-zhang@cauc.edu.cn

*Abstract*—In space-time adaptive processing (STAP) of the airborne radar system, it is very important to realize sparse restoration of the clutter covariance matrix with a small number of samples. In this paper, a clutter suppression method for airborne forward-looking array radar based on joint statistics and structural priority is proposed, which can estimate the clutter covariance matrix in the case of small samples. Assuming that the clutter covariance matrix obeys the inverse Wishart prior distribution, the maximum posterior estimate is obtained by using the low-rank symmetry of the matrix itself. The simulation results based on the radar forward-looking array model show that compared with the traditional covariance matrix estimation method, the proposed method can effectively improve the clutter suppression performance of airborne radar while efficiently calculating.

## I. INTRODUCTION

Space-time adaptive processing (STAP) is an effective tool for clutter suppression and moving target detection in airborne radar systems. The Clutter covariance matrix (CCM) estimation is the critical problem in designing STAP filter. In the classical STAP methods, if we wish to obtain an average loss less than the optimal performance, twice the system degree of freedom (DOFs) independent and identically (i.i.d.) training samples are needed for effective CCM estimation[1]. However, enough i.i.d. training samples can hardly be obtained in real environments.

To improve the STAP performance and reduce the number of training samples, several types of methods have been developed, including the reduced-dimension methods and the reduce-rank methods. The reduced-dimension STAP algorithms utilize data-independent transformations to pre-filter the received signal, and the number of required training samples can be reduced to twice the reduced dimension, some of the methods are auxiliary channel processor (ACP)[2], extended factored approach (EFA)[3], joint domain localized (JDL)[4][5], $\sum\Delta-$STAP [6], and generalized multiple beams (GMB)[7]. The reduced-rank STAP approaches utilize data-dependent transformations, and the number of samples can be reduced to twice the clutter rank. These methods include the orthogonal projection processor (OPP)[8], minimum power eigen canceller (MPE)[8], cross-spectral metric (CSM)[9], and multistage winer filter (MWF)[10].

Recently, a class of algorithms, namely knowledge-aided STAP (KASTAP), has been introduced to exploit additional knowledge of clutter to improve CCM estimation[11]. There are two kinds of prior information are used to improve the performance of CCM estimation. A subset of these techniques concerns the statistical prior information of CCM. A Bayesian approach invoking a complex inverse Wishart prior to the CCM is proposed to estimate CCM in the low-snapshot regime[12]. Other methods try to exploit the particular structure of the CCM to reduce the number of required training samples. The particular structure of the CCM includes persymmetry, Toeplitz property, circulant structure, etc. The maximum likelihood estimation of the persymmetric covariance matrix is discussed, and the number of required measurements can be divided by two. Toeplitz structure is used for covariance estimation in adaptive beamforming and detection[13]. The rank of CCM can be determined by the Brennan rule[14], and a rank-constrained maximum likelihood estimator is introduced for space-time adaptive processing[15].

Our work introduces an advance in robust CCM estimation for airborne STAP radar, known as the low-rank constrained maximum a posteriori estimator. The contributions of this work are formulating and solving a low-rank structured clutter covariance matrix maximum a posteriori estimator for STAP radar, in which no iterative optimization is required and more suitable for real-time processing.

Notations used in this paper are as follows. $A^T$ and $A^H$ denote the matrix transpose and conjugate transpose of $A$, respectively; $\otimes$ denotes the Kronecker product; $\mathbb{R}$ and $\mathbb{C}$ denote the sets of real and complex numbers, respectively; the upper and lower cases boldface letters denote matrices and vectors, respectively; $rank(\cdot)$ denotes the rank, and $tr(\cdot)$ denotes the trace.

The rest of the paper is organized as follows. Section 2 introduces the signal model of a STAP radar. Section 3 exploits the statistical prior information and structural prior information in the clutter. Section 4 presents the proposed

RAM-STAP method and the semi-definite programming (SDP) implication. Section 5 presents the simulation results to demonstrate the performance of the proposed method. Section 6 provides the conclusion.

## II. SIGNAL MODEL

This paper considers a uniformly linear array (ULA) airborne radar, which consists of $M$ antenna elements having a spacing of half of the wavelength ($d = \lambda/2$) and where $N$ pulses are received during the coherent processing interval (CPI) at a constant pulse repetition frequency (PRF) $f_r$. As shown in Fig.1, the platform has an altitude $H$ and moves with a constant velocity $v_p$ along the $x$ axis; $\alpha$ denotes the angle between the clutter patch $P$ and the flight direction; $\beta$ denotes the angle between the clutter patch $P$ and the array line; angles $\theta$ and $\gamma$ denote elevation and azimuth angles, respectively; $\varphi$ denotes the crab angle between the array line and the flight direction, i.e., $\varphi = 0°$ and $\varphi = 90°$ denote the side-looking model and the forward-looking model, respectively.

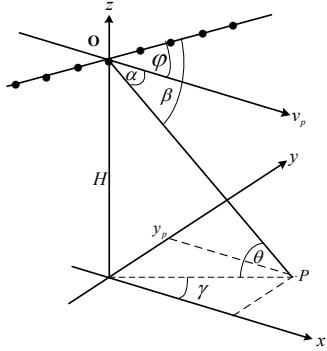

Fig.1. Airborne radar geometry with a ULA antenna

In this work, the range ambiguities are ignored, so the received target-free training sample $x \in \mathbb{C}$ can be expressed as:

$$x = x_c + n \quad (1)$$

where $x$ is an $NM$-dimensional measurement vector, and it is called the space-time snapshot; $n$ is the thermal noise vector; and $x_c$ denotes the clutter vector.

The clutter in a range ring can be modeled as a superposition of signals from $N_c$ independent clutter patches evenly distributed in the azimuth direction, which can be expressed as:

$$x_c = \sum_{i=1}^{N_c} a_i s(f_{d,i}, f_{s,i}) \quad (2)$$

where $N_c$ denotes the number of clutter patches; $a_i$, $f_{d,i}$, and $f_{s,i}$ are the random complex amplitude, the Doppler frequency, and the spatial frequency of the clutter patch; and $s(f_{d,i}, f_{s,i})$ is the $NM$-dimensional space-time steering vector with the Doppler frequency $f_{d,i}$ and the spatial frequency $f_{s,i}$, i.e., $s(f_{d,i}, f_{s,i}) = s_{d,i}(f_{d,i}) \otimes s_{s,i}(f_{s,i})$, which can be expressed as:

$$s_{d,i}(f_{d,i}) = [1 \quad \exp(j2\pi f_{d,i}) \quad \cdots \quad j2\pi(N-1)f_{d,i})]^T \quad (3)$$

$$s_{s,i}(f_{s,i}) = [1 \quad \exp(j2\pi f_{s,i}) \quad \cdots \quad j2\pi(M-1)f_{s,i})]^T \quad (4)$$

where,

$$f_{d,i} = \frac{2v_p}{\lambda f_r} \cos\gamma_i \cos\theta_i \quad (5)$$

$$f_{s,i} = \frac{d}{\lambda} \cos\beta = \cos(\gamma_i - \varphi) \quad (6)$$

The CCM can be defined as

$$R = E[xx^H] \quad (7)$$

Under the zero-mean Gaussian assumption of the thermal noise vector, and according to the principle of the signal-to-clutter-plus-noise ratio (SCNR) maximization, the output of the STAP with adaptive a weight vector $w$ can be expressed as:

$$y = w^H x \quad (8)$$

where the adaptive weight vector is calculated using the CCM as follows:

$$w = \mu R^{-1} s \quad (9)$$

where $\mu$ denotes a nonzero constant and $s$ is the space-time steering vector of the range cell under test (CUT). In practice, the CCM $R$ is unknown and can be estimated from the target-free training samples around the CUT. Assume the clutter of training samples and the clutter in the CUT are i.i.d; then, the CCM can be estimated by:

$$\hat{R} = \frac{1}{L}\sum_{l=1}^{L} x_l x_l^H \quad (10)$$

where $x_l$ denotes a target-free training sample in the $l$th range cell. This is known as the sample matrix inversion (SMI) STAP method[16]. However, the required number of i.i.d. training samples should be twice the DOF to achieve an average performance loss of roughly 3 dB, i.e., $L > 2NM$, which is not feasible in practical applications, especially for the case of a non-side-looking radar.

## III. EXPLOITING CLUTTER KNOWLEDGE

As observed in section 1, a large number of i.i.d training samples are generally not available. Employing the knowledge of the CCM, such as inverse Wishart prior, Teoplitz structures, and sparse structures, can overcome the problem of limited training samples, and improve the performance of CCM estimation. The priori knowledge of the CCM can be classified into two categories: statistical prior information and structural prior information.

## A. The statistical prior information

The CCM can be expressed as a disturbance matrix, and the statistical prior information of the CCM is exploited by the structure of the clutter model or the previous measurements. The clutter model is relevant to the parameter of the radar and platform, such as platform velocity, crabbing angle, terrain data, etc. The CCM is usually characterized by the complex inverse Wishart distribution. For the inverse Wishart prior, the maximum posterior (MAP) estimate is expressed as a weighted sum of a prior matrix $R_0$ and the sample covariance matrix. The probability density function of the CCM is

$$p(R \mid R_0) \propto \frac{|R_0|^v \, \text{etr}\{-(v-N)R^{-1}R_0\}}{|R|^{(v+N)}} \quad (11)$$

Where $\text{etr}\{\cdot\}$ denotes $\exp\{\text{tr}\{\cdot\}\}$, $\text{tr}\{\cdot\}$ is the trace of the matrix, $|\cdot|$ is the determinant of the matrix. The prior matrix $R_0$ is usually obtained by the previous measurements and the echo model of the clutter, such as platform velocity, crabbing angle, etc.

## B. The structural prior information

The structural prior information of the CCM is also be used to reduce the number of required samples.

### 1) Toeplitz-block-Toeplitz structure

When there are no array calibration errors, the CCM in STAP is Toeplitz-block-Toeplitz structured. In a uniformly linear array airborne phased array radar, which consists antenna elements spacing half of the wavelength ($d=$ ) and $N$ pulses are received during a coherent processing interval (CPI), the CCM $R_c$ is a $N \times$ block Toeplizte matrix, defined by

$$R_c = \begin{bmatrix} T & T & \cdots & \\ T & T & \cdots & \\ \vdots & \vdots & \vdots & \vdots \\ T & T & \cdots & \end{bmatrix} \quad (12)$$

$T_i(u)$ is a $M \times$ Toeplizte matrix, defined by

$$T_i(u) = \begin{bmatrix} & & \cdots & \\ & & \cdots & \\ \vdots & \vdots & \vdots & \vdots \\ & & \cdots & \end{bmatrix} \quad (13)$$

### 2) Persymmetric structure

In the STAP radar, we consider a symmetrically spaced linear array to be used for spatial domain processing, and a symmetrically spaced pulse train is used for temporal domain processing[17]. This lead to the disturbance CCM has a persymmetric structure. The persymmetric structure could be exploited to improve to reduce the number of simples.

### 3) Low-rank structure

For STAP radar, the clutter subspace can be spanned by $N_R$ space-time steering vectors, and the clutter covariance matrix can be decomposed as

$$R_c = \lfloor x\, x \rfloor \sum s \otimes s \quad (14)$$

where $x_c = \sum s$ is the clutter signal, $N_R$ is the rank of the $R_c$. The clutter patches are sparse in the angle-Doppler plane, in a uniformly linear array airborne phased array radar, when there are no array calibration errors, the rank of the clutter can be predicted from the system parameters, such as the number of pulses, number of antenna elements, pulse repetition frequency, by Brennan's rule, defined as

$$N_r = \quad R \approx \lfloor \quad \rfloor \quad (15)$$

Where $\beta$ is the Doppler foldover factor, and $\lfloor \cdot \rfloor$ denotes rounding to the nearest smaller integer[18]. $N_r \ll$ , so the CCM is a low-rank matrix.

## IV. LOW-RANK CONSTRAINED STRUCTURED CCM MAP ESTIMATOR

To reduce the number of training samples and estimate the CCM accurately, in this section, we propose a low-rank constrained structured CCM MAP estimator, which utilizes the statistical prior information and structural prior information, such as the low-rank structure, Toeplitz-block-Toeplitz structure and persymmetric structure of the CCM, to estimate the CCM with limited samples accurately.

When i.i.d. multivariate, complex circular Gaussian training samples are available, the **k**th snapshot of the clutter signal is expressed as $x_k \sim \mathcal{CN}\ R)$. $K$ snapshots clutter signal matrix is expressed as $X = [x_1, x_2, \ldots x \quad \mathbb{C}$ , and joint density of $X$ is

$$g(X; R) = \pi^{-np} |R^{-1}|^n \exp\{-tr(R^{-1}XX^H)\} \quad (16)$$

Our goal is to find the likelihood function $g(X; R)$ Maximized positive definite matrix $R$, The logarithm of the likelihood term is:

$$\log g(X; R) = -pn \log(\pi) + n \log |R^{-1}| - \text{tr}(R^{-1}XX^H) \quad (17)$$

For $R$, maximizing the log-likelihood function is equivalent to minimizing:

$$\text{tr}\{R^{-1}XX^H\} - n \log |R^{-1}| \quad (18)$$

Maximum a posteriori (MAP) can be regarded as the statistical prior information of the known covariance matrix based on the maximum likelihood, that is, the probability distribution of the known R. In this paper, we choose the complex center inverse Wishart prior, which is the conjugate prior density of covariance matrix in the multivariate normal model. From $\mathcal{CN}\ \Sigma$ Extracted from $l$ IID samples, $z_k \in \mathbb{C}$ , Then matrix It's a multicenter Wishart,

$\mathcal{CW}\ \Sigma$ .matrix $\mathbf{A}=\mathbf{W}^{-1}$ With inverse Wishart distribution, $\mathcal{CIW}\ \Sigma\ ,l)$ , probability density and $|\mathbf{A}|^{-(l+p)}\exp\{-\mathrm{tr}(\Sigma^{-1}\mathbf{A}^{-1})\}$ proportional.

When combining statistical prior and structural prior for STAP estimation, define an Improved Inverse Wishart Distribution of Nonzero Covariance Matrix on $\mathcal{CIW}\ \mathbf{R}\ ')$ :

$$f(\mathbf{R};\theta)\propto|\mathbf{R}|^{-(l+p)}\exp\{-\mathrm{tr}(\mathbf{R}_0\mathbf{R}^{-1})\} \quad (19)$$

Where $\theta=\{\mathbf{R}_0,\ l,\ r,\ \mathbf{LB},\ \mathbf{UB}\}$ .Then the MAP estimation of R is given as the solution the constrained optimization problem:

$$\arg\max g(\mathbf{X};\mathbf{R})f(\mathbf{R};\theta) \quad (20)$$

where

$$g(\mathbf{X};\mathbf{R})f(\mathbf{R};\theta)= k\left\{\pi^{-pn}\left|\mathbf{R}^{-1}\right|^{n+l+p}\exp\{-\mathrm{tr}(\mathbf{R}^{-1}(\mathbf{XX}^H+\mathbf{R}_0))\}\right\} \quad (21)$$

The maximum a posteriori estimate can be obtained by taking the negative logarithm of the above equation and removing the independent constant:

$$\mathrm{tr}\{\mathbf{R}^{-1}(\mathbf{XX}^H+\mathbf{R}_0)\}-(n+l+p)\log|\mathbf{R}^{-1}| \quad (22)$$

set up $\alpha>0$, $\mathbf{S}=\mathbf{UDU}^H$ is a positive semi-definite matrix, $\mathbf{D}=diag(d_1,\ldots,\ d_p)$, $d_1\geq d_2\geq\ldots\geq d_p\geq 0$, and r is an integer, $1\leq r\leq p$ .then, the solution to the optimization problem

$$\arg\min \mathrm{tr}\{(\mathbf{M}+\sigma^2\mathbf{I})^{-1}\mathbf{S}\}-\alpha\log|(\mathbf{M}+\sigma^2\mathbf{I})^{-1})| \quad (23)$$
s.t. $\mathrm{rank}(\mathbf{M})\leq r$, and $\mathbf{LB}\leq\sigma^2\leq\mathbf{UB}$

Then $(\hat{\mathbf{M}}+\hat{\sigma}^2\mathbf{I})=\mathbf{U\Lambda U}^H$ , when $\mathbf{\Lambda}=diag(\lambda)$ and

$$\lambda_k=\begin{cases}\max(\dfrac{d_k}{\alpha},\mathbf{LB}) & k=1,\cdots\\ \min(\max(\dfrac{\overline{d}}{\alpha},\mathbf{LB}),\mathbf{UB}) & k=r+1,\cdots\end{cases} \quad (24)$$

here, $\overline{d}$ is the average of the characteristic values in the noise space, $\overline{d}=\dfrac{1}{p-r}\sum_{k=r+1}^{p}d_k$ .

The maximum a posteriori probability covariance estimation of $\hat{\mathbf{R}}=\mathbf{M}+\sigma^2\mathbf{I}$ is $\hat{\mathbf{R}}=\mathbf{U}diag(\phi(\mathbf{d}))\mathbf{U}^H$, with $\alpha=n+l+p$ , and eigendecomposition $\mathbf{S}=\mathbf{U}diag(\mathbf{d})\mathbf{U}^H$ for

$$\mathbf{S}=\dfrac{1}{2}\{(\mathbf{XX}^H+\mathbf{R}_0)+\mathbf{T}(\mathbf{XX}^H+\mathbf{R}_0)^T\mathbf{T}\} \quad (25)$$

that $\dfrac{1}{2}\{(\mathbf{XX}^H+\mathbf{R}_0)+\mathbf{T}(\mathbf{XX}^H+\mathbf{R}_0)^T\mathbf{T}\}$ is the projection of the sample covariance matrix onto the closed, convex set of T-conjugate symmetry matrices. The symmetry may be exploited for computational efficiency in computing STAP weight vectors using the estimated covariance, $\hat{\mathbf{R}}$.

## V. NUMERICAL RESULTS

To verify the effectiveness of the method in this paper, the simulation data is generated through the following parameters: the antenna array is the number of elements $M=8$ uniform linear array, element spacing $d=\lambda/2$, radar operating wavelength $\lambda=0.66\,\mathrm{m}$ , Number of coherent pulses $N=8$, the height of carrier platform $H=9000\,\mathrm{m}$, clutter detection range $90000\,\mathrm{m}$, carrier speed $v_p=50\,\mathrm{m/s}$, pulse repetition rate $f_r=300\,\mathrm{Hz}$ , 360 clutter units in $-180°\sim$ uniform distribution, distance resolution is $37.5\,\mathrm{m}$, the experiment compares the maximum a posteriori of the low-rank constraint and symmetric constraint proposed in this paper（rank and symmetric constraint maximum a posteriori，RS-MAP）method, SMI method and maximum a posteriori estimation with incomplete use of prior information（maximum a posteriori，MAP）、rank constrained maximum a posteriori(R-MAP) and symmetric constrained maximum a posteriori(S-MAP）method.

### A. Clutter space-time power spectrum analysis

Experiment 1: Compare the influence of different prior information on sparse recovery STAP under fewer snapshots. The array is erected with a forward-looking array, and 8 snapshot data are used for sparse recovery. The clutter rank is set to 20. Because it is difficult to accurately estimate the clutter rank in the actual clutter environment, and there is no clear analytical expression so far, the selection of the rank in this paper is only an empirical value expanded according to Brenna's criterion, and auxiliary data is generated through 8 snapshots in advance, The prior matrix R0 is obtained through the clutter covariance matrix inversion algorithm as the known clutter statistical prior information. Figure 2 shows the space-time spectrum of clutter sparsely recovered by different methods for airborne forward-looking radar.

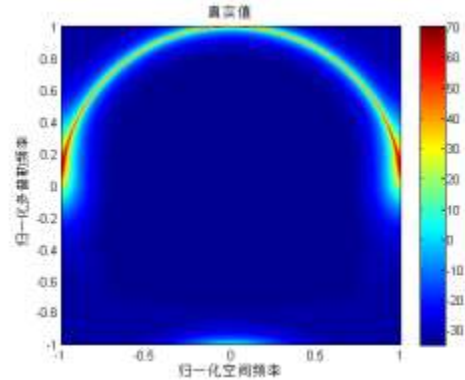

(a) Real clutter power spectrum

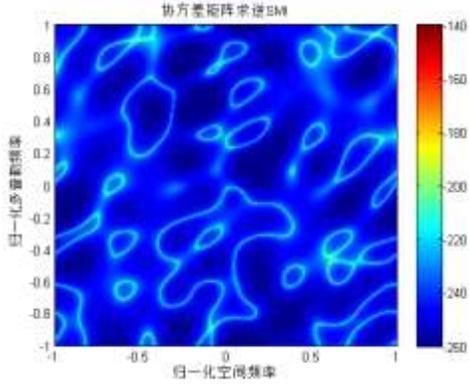
(b)SMI method

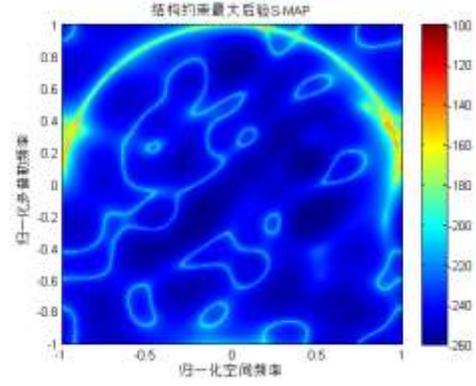
(e)S-MAP method

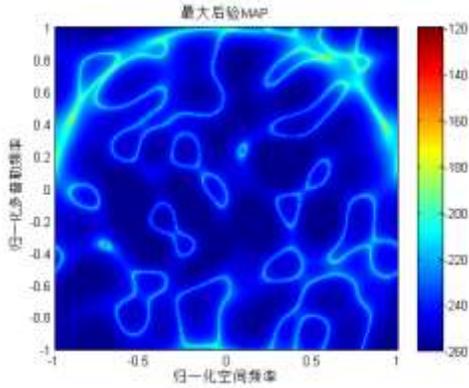
(c)MAP method

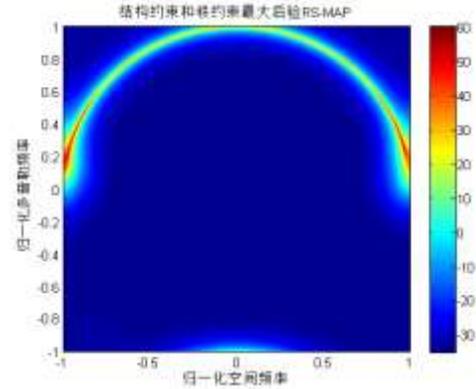
(f)RS-MAP method

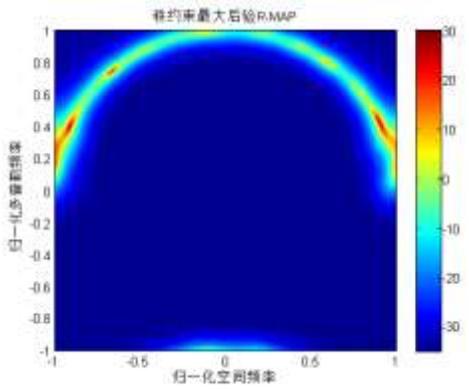
(d)R-MAP method

Figure 2 Spatial-temporal spectrum of clutter sparsely recovered by different methods for airborne forward-looking array radar

It can be seen from Figure 2 (b) that due to the serious shortage of IID samples required by the SMI method, the estimation result of the covariance matrix is very poor, and even no clear clutter ridge can be seen in the power spectrum. It can be seen from Figure 2 (c) that after adding the statistical prior information that the clutter covariance matrix obeys the inverse Wishart distribution, the MAP method is significantly improved compared with the SMI method, but the estimation is still inaccurate. In Figure 2 (d) and (e), rank constraint and structural symmetry are added based on maximum a posteriori, respectively, and the clutter space-time spectrum obtained is much better than the traditional SMI method. In particular, Figure 2 (f) RS-MAP is a clutter suppression method proposed in this paper based on joint statistics and structural a priori. By using the low-rank matrix recovery theory and adding inverse Wishart distribution and symmetry characteristics, it can achieve more accurate clutter recovery under the condition of small samples. It can be seen from Figure 2 that under the same snapshot conditions, the more prior information is added to the sparse recovery method, the better the recovery effect.

## B. Clutter suppression performance analysis

Experiment 2: Figure 3 shows the SNR loss curve of airborne forward-looking array radar obtained by 1000 Monet Carlo experiments under 16 snapshots. It can be seen from the figure that the more prior information is added to SMI, MAP, R-MAP, S-MAP, and RS-MAP, the narrower the null is, and the closer it is to the true value. In addition, under the condition of 16 snapshot data, SMI, MAP, and P-MAP have a large loss of signal-to-noise ratio, which will greatly reduce the effect of the system on clutter suppression. Under the same conditions, the RP-MAP method proposed in this paper, which combines statistical prior information and structural prior information, can use a small amount of snapshot data to sparse recover the clutter covariance matrix and can form a narrow null in the main clutter area, thus achieving more accurate clutter suppression.

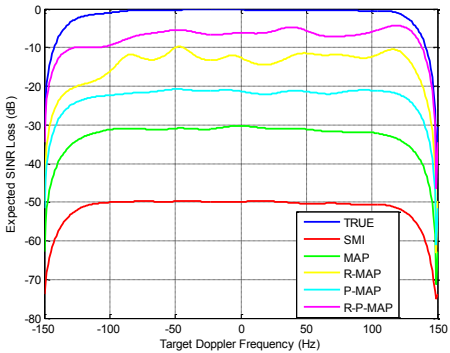

Fig. 3 Signal-to-noise ratio loss curve for 16 snapshots of different methods

Experiment 3: To better compare the clutter suppression performance of the methods with different prior information under different snapshot numbers, Figure 4 draws the curve of the average SNR loss of the methods with different prior information varying with the number of snapshots, where the number of snapshots varies with 8:8:96, and the number of Monet Carlo experiments is 10000.

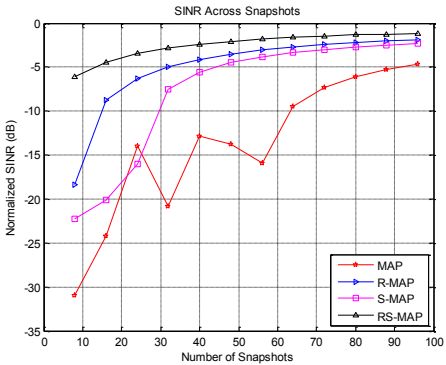

Fig. 4 Curve of average SNR loss versus snapshots

It can be seen from Figure 4 that when the number of snapshots is 8, the average signal-to-noise ratio loss of RS-MAP is nearly 25 dB higher than that of MAP, and the average signal-to-noise ratio loss of R-MAP with prior information added is also about 5 dB higher than that of S-MAP. Moreover, when the signal-to-noise ratio loss is - 5dB, the number of snapshots required by the RS-MAP method is reduced by about 75 and the number of samples required by the MAP method is reduced by six times. In the case of a small number of samples, the performance of RS-MAP and R-MAP is stable and the estimation effect is good, but the MAP method fluctuates greatly and the estimation performance is inaccurate. As the number of snapshots increases, the estimation performance of MAP shows an upward trend, which is close to the estimation performance of the other three methods with prior information added. However, it can be seen that the performance of the joint statistics and structure prior RS-MAP method has always been better than the other three methods.

## VI. CONCLUSION

Aiming at the erection of airborne forward-looking array radar, this paper uses the sparse recovery method under the condition of small samples to achieve clutter suppression in space-time adaptive processing and proposes a clutter suppression method that adds prior information based on low-rank matrix recovery. This method not only uses the sparse characteristics of target echo in the angle Doppler domain but also uses the statistical distribution and symmetry characteristics of the clutter covariance matrix itself, The sparse restoration of the clutter covariance matrix is realized according to the low-rank matrix restoration theory. Simulation results show that this method can effectively recover the clutter covariance matrix and improve the clutter suppression performance when there are only a few independent identically distributed samples in the case of airborne forward-looking array radar.